# An Asymmetric Sparse Telescope

B. Martin Levine, Michael Kaplun, And Erez N. Ribak*
* eribak@physics.technion.ac.il
*Physics Department and Asher Space Research Institute, Technion – Israel Institute of Technology, Haifa 32000, Israel*

**Abstract:** We designed and built a novel model of a deployed space telescope which can reliably align its segments to achieve the finest possible resolution. An asymmetric design of both the segment shapes and their pupil locations were tested in simulation and experiment. We optimised the sparse aperture for better spatial frequency coverage and for smoother images with less artifacts. The unique segment shapes allow for an easier identification and alignment, and the feedback is based only upon the focal image. The autonomous alignment and fine tuning are governed by mechanical simplicity and reliability.

## 1. Introduction

Both ground and space telescope are strongly constrained in volume and mass. The solution to the size limitation in both cases can be telescopes made up from thinner mirror segments which combine to form a large aperture primary. Generally, these segmented telescopes keep their sectors very close to each other, avoiding background radiation from behind the gaps while maximizing the collection area. On the other extreme, interferometers gain in resolution by increasing the size of the aperture and by making it sparse, even extremely sparse (in the radio regime), all at the price of signal lost between the elements.

While it is not easy to perfectly align any telescope, the task becomes tougher when the telescope is segmented and sparse [1]. Instead of worrying about the limited degrees of freedom (DoFs) of the few elements, now each added segment contributes its own DoFs, anywhere between one and six. That is even while excluding optional sub-segment correction (such as curvature adjustment or surface errors). The wave-front accuracy required is that of a contiguous telescope, traditionally a small fraction of the shortest wave-length for observation. Finally, if the telescope is remote (on the ground or in space), the alignment has to be done without human intervention.

A large body of methods has already been devised to solve these issues, and to measure the individual wave-front deviations. They are roughly divided between aperture plane and focal plane (wave-front) sensing (WFS). Aperture plane WFS usually entails using additional sensors or filters to measure the wave-front derivatives of the segmented telescope or interferometer sub-apertures (Section 4). Additional sensors allow proximity measurements of segment-to-segment error, albeit limited to the edge accuracy [2, 3]. Unfortunately, they are less practical for sparse apertures with large gaps.

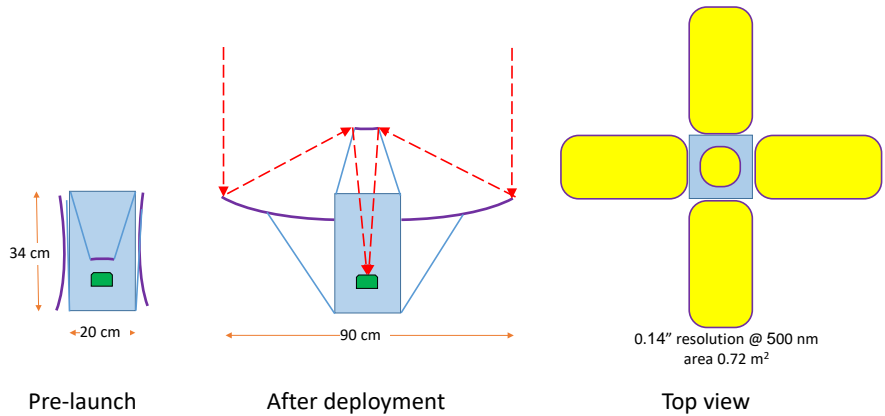

Fig. 1. Concept of a basic deployable telescope.

Focal plane WFS does not require additional hardware, a significant advantage in autonomous telescopes, where simplicity reduces the chance of equipment failure. It entails measuring the focal intensity distribution, which is non-linearly related to the aperture amplitude and phase. The notion is that less aberrated wave-fronts produce images closer to the diffraction limit, namely sharper images, by Parseval's theorem [4, 5]. How each DoF affects the image sharpness may not be directly known, so optimisation in hardware seems a straightforward way to achieve wave-front control. On smaller systems this optimisation is very easy, but for segmented telescopes with many DoFs, the search volume becomes overly large and unfathomable. Using focal image optimization methods such as simulated annealing [6] is safe and slow. Unfortunately, when combined with faster steepest descent [7, 8] the search might sink into sub-optimal solutions. It was proposed to use a supervised learning algorithm [9], where a full search must first be performed over the same huge exploration volume ahead of time, and memorized for a one-time application. Phase diversity WFS iteratively uses a focal image and an extra-focal one, and works best for compact, narrow-band objects, and contiguous apertures [9]. The solutions that we propose below pertain to segmented and sparse space telescopes. In some cases, they can be applied to ground-based telescopes.

## 2. Aperture Design

We concentrate here on space telescopes for plain imaging, such as earth surveys or imaging stellar fields. Further constraints on the design can also be added (Section 4). Fig. 1 shows us the basic idea of a small space telescope in the Fresnel configuration, which is simple to construct and deploy [11, 12]. It has two drawbacks: the segments' positions and shapes. First, the segments cannot be measured with an aperture WFS, because of the large gaps between them. Stellar interferometry had shown that the focal-plane fringes between every two segments in the aperture must be unique in order to identify them. These two segments contribute to two symmetric and distinct lobes ("base-line") in the complex Optical Transfer Function (OTF) or its modulus (MTF). The OTF is the autocorrelation of the pupil map, and is obtained by Fourier transforming the Point Spread Function (PSF) – the image of a point source. However, each coordinate in the MTF map is the sum of all equal base-lines (same spatial frequency). In other words, to distinguish each such pair of segments, these base-lines must somehow be made identifiable or otherwise non-redundant [13-18]. The inclusion of more, redundant base-lines does not contribute to the resolution, but it improves the signal-to-noise ratio (important for some cases). Current research on further aspects of sparse aperture imaging deals with identical and symmetrical segments, such as circles and hexagons [16-19].

To break segment symmetry, we devised an optimized search program, to place the segments around a central hole, as in a Cassegrain set-up (Fig. 2). These sectors were cut out of a parabolic mirror, and for mechanical simplicity we used only four segments. Notice that now the base-lines 1-2 and 3-4 are distinct, as opposed to the concept in Fig. 1. The MTF (Fig. 2e) serves as the deconvolution kernel for images obtained with the segmented sparse telescope [20].

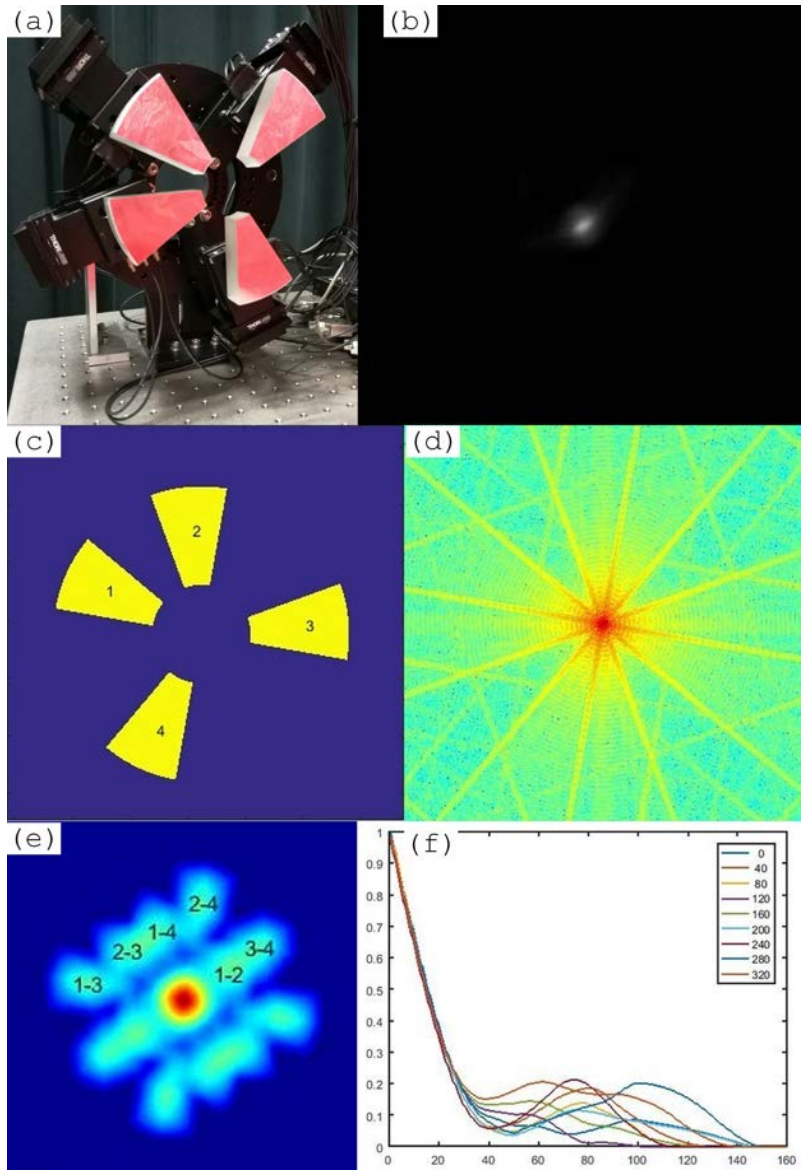

Fig. 2. (a) A model of a 20 cm sparse telescope (reflecting off a red colored sheet); (b) The PSF of a single segment has two spokes and a unique orientation; (c) The sectors are set at non-redundant angles, rotated with respect to (a); (d), The calculated PSF, with strong scattering off the straight sector edges (log scale); (e) The central part of the MTF, with limited gaps between the lobes; (f) Radial cuts in the MTF for various orientations, where the mid-frequencies can drop to zero.

Here we see the other main problem with the design in Figs. 1 and 2: the long straight edges of the segments diffract (scatter) a great deal of light, evident as radial spokes in the PSF (Fig. 2b, 2d), and inhibit the detection of faint objects near a bright one. Of course, the optimal shapes are those with the best area-to-perimeter ratios, i. e. circles, but these cannot be distinguished as they have the same diffraction pattern (assuming similar or equal sizes). Thus, to minimize diffraction, we tried different shapes of rounded segments, from circular to elliptical to egg shapes (ovoids). Ellipses are also advantageous from the manufacturing point of view, even more so if they have the same shape, and for mechanical and thermal stability of the telescope. Ovoids [21] provide more reflective surface compared to ellipses, closer to the sector shapes but more rounded. For example, a typical ovoid can be given by a deformed ellipse, $(x/a)^2 + e^{2x}(y/b)^2 \leq 1$, where $a \approx 1/4$, $b \approx 4$. However, for our proof-of-principle, ovoids are slightly more complex in calculation, so we used ellipses for this study.

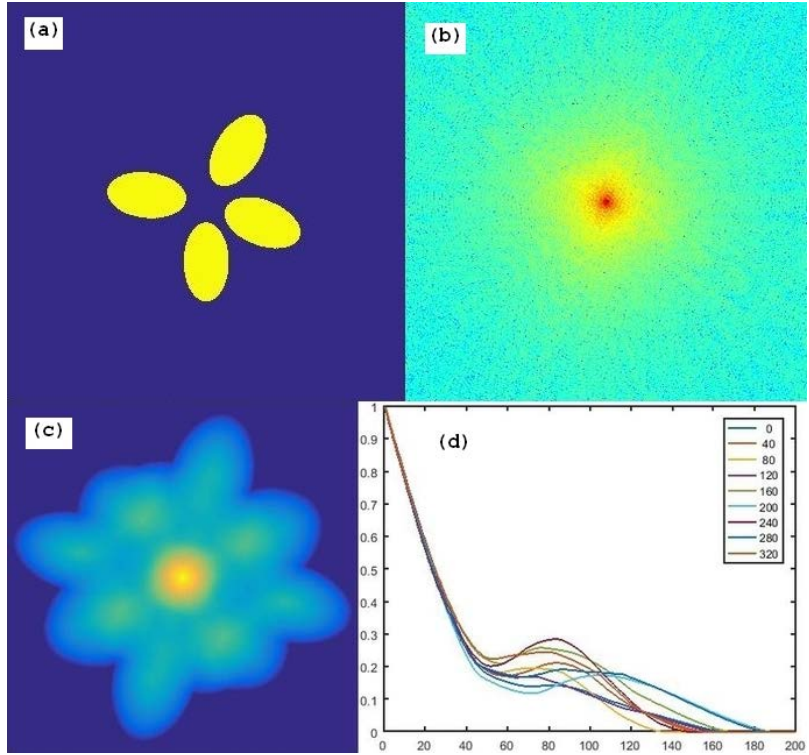

Fig. 3. (a) Elliptical segments, in a typical realization, to be compared with Fig. 2; (b) The PSF (log scale); (c) The central part of the MTF. No gaps exist between the OTF lobes; (d) Radial cuts in the MTF, with good mid-frequencies response.

The segments can be placed non-redundantly anywhere in the aperture plane [16-19]. For deployment simplicity we reduced the search space, and placed the ellipses around a single central circle. Each ellipse $\varepsilon_i$, of semi-axes $a$ and $b$, is radially shifted by $ka$, then rotated about the telescope center by an angle $\alpha_i$ (Fig. 3),

$$(x'\cos\alpha + y\sin\alpha)^2/a^2 + (x'\sin\alpha - y\cos\alpha)^2/b^2 \leq 1; \quad x' \equiv x - ka. \tag{1}$$

The pupil function is the sum of ellipses, $P = \Sigma_{i=1,..4}\,\varepsilon_i$, where the ellipses $\varepsilon_i$ are pairwise disjoint. The pupil auto-correlation (or MTF) is $Q = P \otimes P$ with a standard deviation $S = \text{STD}(Q)$. The area of $Q$ (where $Q \gtrsim 0.01$) is $A$. If the telescope diameter is set, the optical throughput also grows with $A$. We ran an automated optimization with the axes ratio $a/b$ of the four segments as one free parameter. Three more parameters were their angular separations $\alpha_i - \alpha_j$ around the telescope aperture. There were two cost functions that we could have used for optimization. The first target criterion was the widest MTF (maximal MTF area $A$), namely with the broadest $u$-$v$ (Fourier frequencies) coverage. As the MTF can be non-contiguous or clumpy (Figs. 2e, 3c), we tried another optimisation, that is to reduce it (minimal standard deviation $S$). The two

criteria – the widest MTF or the smoothest (least clumpy) MTF - were very similar, and so were their results.

In many cases, narrow gaps sometimes crept among the MTF lobes, and these solutions had to be excluded. This is important, since the images obtained through the sparse telescope are to be deconvolved by this MTF to enhance their weaker spatial frequencies. Thus, the MTF should not be close to zero, to avoid small-number division in Wiener filtering of the images [20].

We ran many trials with multiple solutions, which we sorted to correct for the rotation and reflection symmetries (Fig. 3). All configurations had figures of merit, either MTF area or MTF smoothness, which varied among them only by a few percent. However, the ellipticity of the segments always converged to be either maximal or minimal, according to either optimization criterion. Since launch packaging constraints will determine the final ellipticity (or egg shape), we set first the axes ratio, then optimized only the angles among the segments. For off-center shift of the segments (Eq. 1) $ka = 1.5a$ and axes ratio $a/b$ of 1.5 to 2.2, the azimuthal angles between segments came out to be 66.8º±3.4º, 83.8º±5.2º, 113.0º±4.5º and 96.4º±6.3º. The rounder edges produced a more regular PSF than straight ones (compare Figs. 2f and 3d). Numerically, the trends were weak and opposite: for the same $b/a$ range of 0.45 to 0.67, the autocorrelation area $A$ rose linearly from 0.61 to 0.69, relative to the full aperture area. The MTF smoothness $S$ had the same weak behaviour, dropping from 0.122 to 0.112. This means that broader autocorrelation was also lumpier, since longer ellipses produced more serrated MTF edges.

## 3. Experiment

To test our ideas of the aperture design and its basic consequences, we used our laboratory model as a proof of concept. That setup also served to test issues arising from inexpensive actuators and lower quality optics: During construction of the experiment, we realized that the reflective surfaces close to the straight edges of the segments were turned up or down, as a result of stress release in the glass during segmentation. To reduce the edge effects, and also to round down the segments, we put ovoid masks in front of them. This significantly reduced the spokes in the PSF, as well as created smoother MTF of the aperture (Fig. 4).

The unique aperture design led to a different treatment of two alignment issues, correcting tip-tilt errors of the segments, and for their piston phasing (correction for optical path differences). Previous studies mostly assumed perfect mechanical control, down to a few nanometer feedbacks, which we try to avoid in our simplified design for both laboratory and space actuators. From previous experience [22, 23], and from our own experience, we knew that correcting for tip-tilt is the easier problem: We have used simulated annealing [5-7] to optimize the PSF by maximizing the image sharpness. We have found that for segmented telescopes, correction for tip-tilt was fast, but phasing was very slow, essentially an asymptotic convergence process. With the new approach, we started with the initial PSFs of the separate segments, caused by their focus errors. (It is assumed that the camera field-of-view is large enough to capture all four foci from the start.) We were able to bring the segments into focus with ±0.5 mm optical path difference error (or focal depth). Next, we modeled and created a control matrix for the tip-tilt motors. These were piezoelectric actuators (Thorlabs KC1-PZ_M) with nanometer steps (range ±4 µm), driven by high voltage electronics (adapted from a Wavescope®, Adaptive Optics Associates). These PZT actuators were connected in series with coarser screw actuators for focus scan (Thorlabs Z812B, range ±6 mm, 29 nm steps). The light source was a single-mode fibre-coupled white LED (Prizmatix W3000M nm) at the $f$/6 focus of a 200 mm collimating mirror.

During modeling, the system checked the impact of every motor on the position of the segment PSF, and using this information built a control matrix. Additionally, it checked the orientation of the segment PSF arising from the broken rotational symmetry of the ovoid apertures [23]. The program did not assume known segment orientations or shapes. Instead, it learned them during this stage from the binarized PSFs of the focused segments (Fig. 5), which were nearly elliptical. The major axis of each PSF was first identified and saved, then compared with the orientation of each one of the stored PSFs for their identification. This identification method is robust even when these PSFs merge and overlap after stacking. It replaces the method of tilting out and then tilting in each segment to identify it [22, 23]. Hysteresis in the piezoelectric actuators also made the tilt method less reliable.

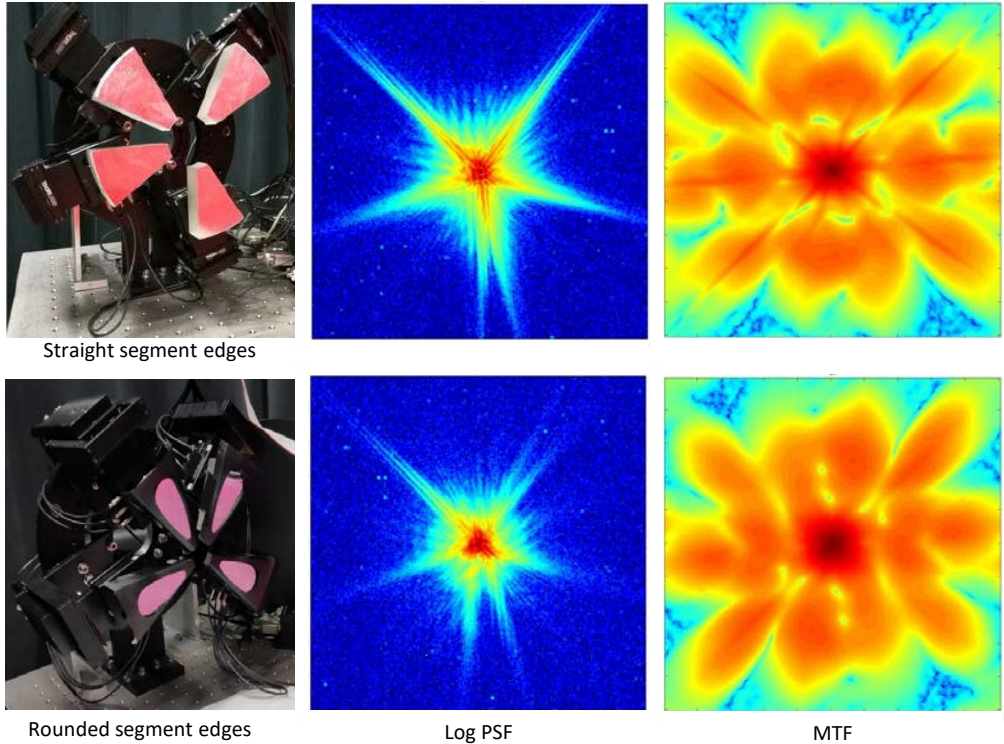

Fig. 4. The straight segment edges (left, top) created a spiked PSF (center) and MTF (right). By masking the segments' edges (bottom row) the spikes were reduced and the MTF smoothed. These masks also hid the low-quality surface errors at the periphery of these segments. (The segments orientations were also changed).

The PSFs positions in the image plane are proportional to the tip-tilt errors of the segments. These errors were corrected via the control matrix, corresponding to each piezoelectric motor. This accomplished the "stacking", the angular alignment of the segments' PSFs.

After the tilt on the segments was minimized, the last and longest step was to phase ("piston") them to be at the same distance from focus, within a fraction of a wave-length. We applied a method akin to optical coherence tomography (OCT): we scanned one segment and observed the combined PSF of all segments, to see when it varied in intensity [1, 24]. Such a variation meant that fringes ran through the image in focus (Visualization 1). Only at that time, new lobes appeared in the MTF (see below). The white light spectrum limited the coherence length of the beam very much, so fringes were only visible for a few wave-lengths before and after they had equal paths among the segments (approximate range ±3 µm, Fig. 6d). Although narrow flip-in spectral filters could increase this range, we sought to eliminate additional mechanical components, so as to maintain system simplicity in space. For earth-observing and for astronomical telescopes we assumed using an unresolved star for the initial alignment stages. We have pre-viously aligned successfully a 7-DoF system on an extended ground scene in the lab by maximizing its contrast. However, optimisation of such a scene during a short fly-over could be difficult [11].

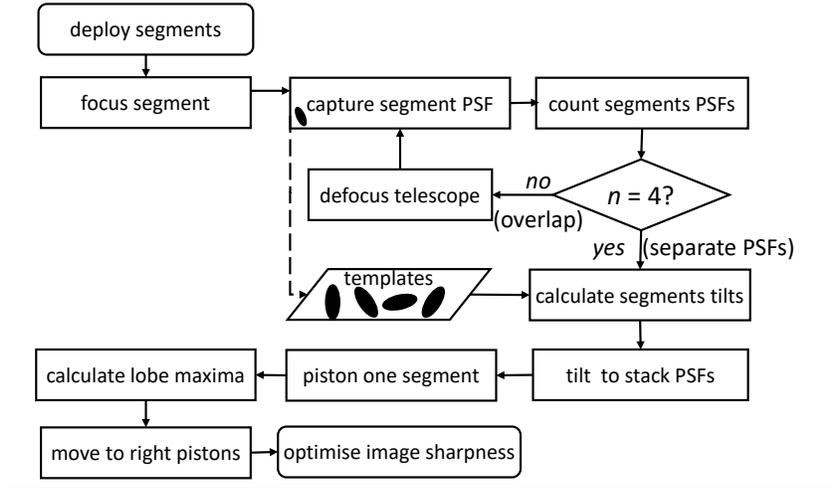

Fig. 5. Flow-chart of the process of correcting for tip-tilt and then piston of the telescope segments.

Since the segments were placed at non-redundant positions (Fig. 4), it was possible to identify the two segments that contributed to the fringes at each specific lobe. First, we moved segment 4 by 500 nm steps, until we could see fringes crossing the MTF lobe corresponding to segment 4-1, 4-2, and 4-3 (Fig. 6a). During each crossing, the signal at the lobe (as in Fig. 2e) would increase until reaching a threshold. At that point the step size would switch to finer 28.5 nm steps (Fig. 6b), in order to find a maximum during phasing; when the pair lost phasing, the step size would return to 500 nm. These fringe crossings happened each at a different scan position, and provided the optical path differences with the other three segments. During very long scanning, we had to occasionally maintain the tip-tilt to keep the PSFs overlap, with the fringes on top. There is also the option of scanning all four segments against all others (assuming a perfect closed loop actuation), in which case a least-squares solution could provide the relative piston values [25]. No use was made of the constraint of zero phase closure [26], which could improve performance even further, especially when dealing with ground-based telescopes.

After the scan was finished, one scanning segment, segment 4 in our experiment, became the stationary reference, and the others were moved to reach phasing with it. We did not want to rely on the repeatability of the pistons, or on the possible difference in mechanical conditions between the pistons. Hence, after bringing the pistons of segments 1-3 to the position (recorded from the previous scan), a new short-range single segment scan began. Again, the fourth segment beam interfered with all other segments, resulting in simultaneous fringe scanning in all MTF lobes. After this final scan all segments were at most within ~50 nm from each other. In Visualization 1 one can see the development of the MTF during a scan for segment 4, where the previous three are almost phased. It is possible to see rise of the lobes 4-1, 4-2, and 4-3 (marked) when the scanned segment gets near and then crosses the equal path point.

PSF images were collected by a 5472×3648, 2.4μm pixel camera (Basler acA5473). Each image was Fourier transformed instantly to obtain the complex OTF, and sampled at the center pixels of the non-redundant OTF lobes (Fig 4., right). Some loss in accuracy is due to the low quality of the segments. While the local surface smoothness was ~10 nm, the masked area had ~60 nm surface standard deviation due to low-frequency bending of the edges. In addition, more area of all twelve OTF lobes could be used to better determine the fringes, except where they overlap with the next OTF lobe. Thus, the non-redundant masking demands could be relaxed in some cases: the relative segments sizes can be increased, enlarging also the overlap area of the side lobes, as long as we have at least one point in each OTF lobe which is unique to its pair of segments only [15, 18].

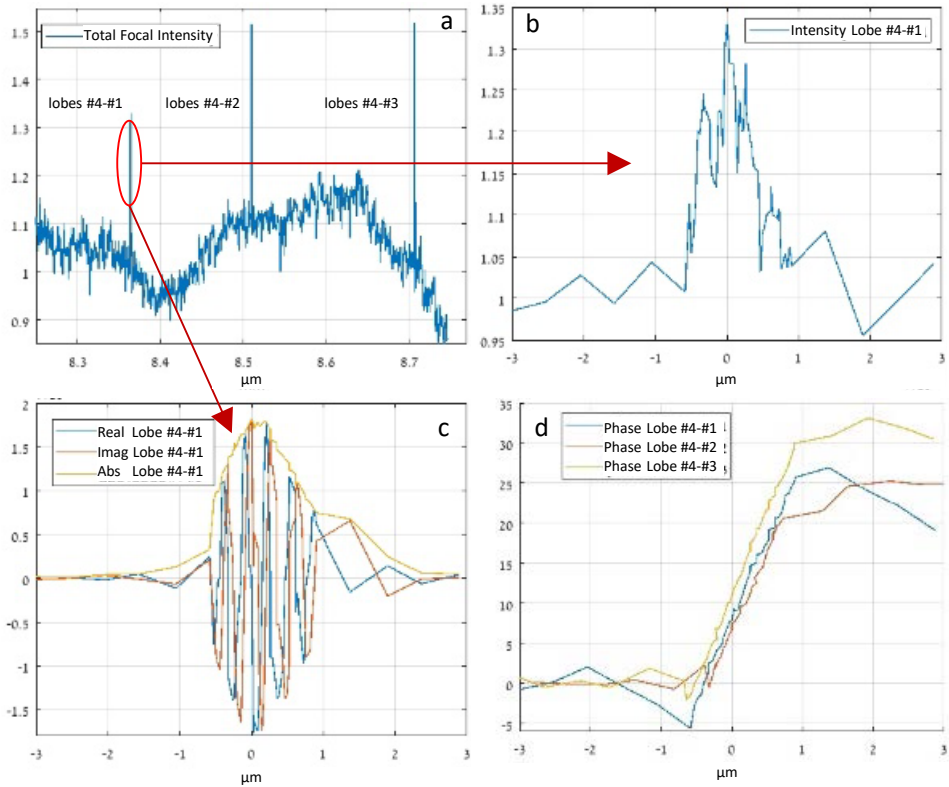

Fig. 6. (a) Scanning for mutual fringes between segment #4 and the other three, by looking at the PSF intensity. (b) Magnified intensity of the first encounter. The intensity enhancement is about 30%. (c) Instead, it is possible to look at the Fourier transform of the PSF (i.e. the OTF, Fig. 4, right), and see which of the complex lobes rises from zero, and reaches its maximum amplitude (Visualization 1). (d) Cumulative phase during the different scans against all other lobes (radians).

The whole alignment process (Fig. 5) is to be compared with the optimization alternative [7]: gradient descent, simulated annealing, and stochastic parallel gradient descent. In a laboratory setup, there are at least 17 DoFs (tip, tilt, fine and coarse piston for each segment, and global focus), where each of these has about 1,000 gradations. This makes the full search volume for general optimization methods enormous, $\sim 10^{51}$ options. But when starting near the best solution, and dropping the DoFs of coarse pistons and global focus, we get down to 13 DoFs, and 100 gradations each, namely $\sim 10^{26}$ options. With the chances of multiple minima considerably reduced, gradient search suffices. Because of the low quality of the segments, even after masking their edges, we could not reach the theoretical diffraction limit. For that reason, we also did not try to optimize the secondary mirror or the camera with this setup.

The time limitation for the full alignment on orbit is quite weak: a new space entity requires days to set itself up. Still, it is beneficial to intelligently limit the search volume, which is what we propose and demonstrate here. Focusing and segment alignment can take minutes, and scanning one segment against the others, while searching for mutual fringes, can take tens of minutes: the range is a few millimeters, and the step size – a fraction of a micron. But by the time these two main steps are over, the segments are very close to their optimal position, and search is faster. All our runs (including constructing the control model) were less than six hours.

Future development, to achieve nanometer-scale phase deviations, can be reached by several methods [22, 25-30], some of which require accurate actuators, narrow band light, removable masks, or smaller initial errors. Without these, and with possible open loop commands, one could resort to PSF sharpness optimization to achieve that position. Notice that these search procedures are not limiting: if they fail (for example by a malfunctioning actuator), simulated annealing can still optimize the image sharpness, from initial position, with the constraint on the bad component included indirectly.

When using simulated annealing optimization, calibrated step sizes or fully decoupled motions are not important, as opposed to gradient search and phase diversity methods. (The only exception is the focus motor, which accurate position is required for later operation of the telescope, e.g. to correct for thermal distortions of the structure). Erroneous assumptions about the locations of the motors or actuators are more important in the second stage, of phasing the elements, since long motions, with attendant larger errors, are unavoidable. However, the search scheme (Fig. 5) can be repeated if the image quality is insufficient, before starting the final optimization stage.

## 4. Conclusions

The design of a generic telescope (for earth or astronomical imaging) drives some of the critical constraints in this study. The initial incentive for our search for a new design was the need for a co-alignment and co-phasing of a simple telescope under these constraints. But this can be taken further: we can think of enhancements for special cases. For example, it is clear that a sparse telescope will have severe thermal background if operated in the infra-red. In that case, it would require a cold shield to mask off the gaps among the segments. A science-driven example is extra-solar planet imaging. In this case, the PSF can be optimized such that it is blanked out near the main star, where any planets, if they exist, stand out clearer. Furthermore, rotation of the telescope on its optical axis might help remove instrumental artifacts [24]. Of course, these drivers require a totally different design of the segment shapes and of the occultation mask, which is beyond the scope of this study.

To summarize, we demonstrated here a novel method to shape, place and align sparse telescope segments. We have optimised and then used both the non-redundant shape of the aperture segments, and the non-redundant positioning of the segments, to tell them apart, and to identify the segment pairs, based on previous simulation. The search volume for the best segment alignment was significantly reduced, and with it the total alignment process time. All of these were achieved with a minimally complex laboratory model of a space telescope with no additional hardware, and with minimal assumptions on mechanical or spectral precision.

**Acknowledgement**. The authors thank Peter Tuthill for suggesting other aperture designs.

**Disclosures.** The authors declare no conflicts of interest.

**Data availability.** Data presented in this paper may be obtained from the authors upon reasonable request.